# The Remixing Dilemma: The Trade-off Between Generativity and Originality

Benjamin Mako Hill (mako@mit.edu)
Andrés Monroy-Hernández (andresmh@media.mit.edu)

In this paper we argue that there is a trade-off between generativity and originality in online communities that support open collaboration. We build on foundational theoretical work in peer production to formulate and test a series of hypotheses suggesting that the generativity of creative works is associated with moderate complexity, prominent authors, and cumulativeness. We also formulate and test three hypotheses that these qualities are associated with decreased originality in resulting derivatives. Our analysis uses a rich data set from the Scratch Online Community – a large website where young people openly share and remix animations and video games. We discuss the implications of this trade-off for the design of peer production systems that support amateur creativity.

## I  INTRODUCTION

Remixing, the reworking and recombination of existing creative artifacts, represents an enormous, important, and controversial form of online engagement. Most commonly used in reference to the creation of music, video, and interactive media, Manovich (2005) has called remixing, "a built-in feature of the digital networked media universe." Lessig (2008) has argued that remixing reflects both a broad cultural shift spurred by the Internet and a source of enormous creative potential. Benkler (2006) has placed remixing at the core of "peer production" – the organizational form behind free and open source software and articles on Wikipedia – and has argued for the deep cultural importance of remixing. Scholars of innovation have suggested that remixing practice plays a critical, if under-appreciated, role in new forms of innovation facilitated by the Internet (e.g., von Hippel, 2005).

To advocates of remixing, and to proponents of peer production more generally (e.g., Raymond, 1999; von Krogh and von Hippel, 2006), the fecundity or "generativity" of creative works is of utmost importance in that it determines remixing's very existence. But although collaboration lies at the heart of definitions of peer production, and despite the enormous amount of collaboration that occurs in some of its poster children, most articles on Wikipedia and other wikis never attract many editors (Ortega, 2009; Kittur and Kraut, 2010), most Free and Open Source Software (FLOSS) projects founder (Healy and Schussman, 2003), the majority of YouTube videos are never remixed, and most attempts at "meme spreading" on 4chan fall flat (Bernstein et al., 2011).

Proponents of remixing argue that generativity leads to increased innovation and democratized production. For example, Zittrain (2008) argues that some technologies, like the Internet, are generative and important not because they solve problems directly but because they provide rich and unconstrained platforms upon which derivative technologies can be built. Previous research has tangentially looked at ways to promote remixing, especially among young people (e.g. Cheliotis and Yew, 2009; Luther et al., 2010; Hill et al., 2010). Jenkins et al. (2006) have argued that educators can work to increase remixing behavior in young people.

Although remixes are defined by their derivative nature, the promise of remixing is contingent on the *originality* of derivative works. Remixing in the form of near-perfect copying seems unlikely to achieve Benkler's (2006) goal of "making this culture our own," or in building the transformative and empowering improvements at the heart of Zittrain's examples. We also know users of remixing communities react negatively to visibly similar remixes of their projects (Hill et al., 2010). Moreover, issues of originality are often at the center of moral and legal discussions of remixing (Aufderheide and Jaszi, 2011). For example, to receive protection under US copyright law, a derivative work must be original in the sense that it is, "independently created by the author (as opposed to copied from other works), and that it possesses at least some minimal degree of creativity" (O'Connor, 1991). A critic of peer production, Keen (2007) conflates remixing on the Internet with copying saying, "the pasting, remixing, mashing, borrowing, copying – the stealing – of intellectual property has become the single most pervasive activity on the Internet."

In this paper, we look at behavior within a peer production community to try to understand how designers of peer-production systems might, or might not, be able to support remixing that is both *generative* (i.e., likely to engender derivative works) and *original* (i.e., derivative works differ substantially from their antecedents). We attempt to answer two related research questions. First, what makes some creative works more generative than others? Second, what makes some creative works engender more transformative derivatives? Previous descriptive and theoretical work on peer production has pointed toward answers to both questions but has largely eschewed testable theories and hypotheses. Elaborating on foundational theory in peer production, and supported by empirical tests, we suggest that the answers to these two questions point in opposite directions and imply a trade-off for designers seeking to support remixing in online communities. *We suggest that three factors associated with higher levels of generativity – moderate complexity, creator prominence, and cumulativeness – are also associated with decreased originality in the resulting remixes.*

Using data from Scratch – a large online remixing community where young people build, share, and collaborate on interactive animations and video games – we present evidence that supports and extends several widely held theories about the foundations of generativity and originality. Our results suggest that designers of online collaborative communities may face a dilemma obscured by those celebrated exemplars of peer production communities: that system designs that encourage and support increased rates of remixing may also result in more superficial products.

In Section II of this paper, we discuss theoretical scholarship on remixing and motivate a series of six hypotheses about the predictors of generativity and originality, each of which we state explicitly following a description of relevant literature. In Section III, we present our empirical setting and analytic strategy for testing these hypotheses. In Section IV we present the results of our analysis and in Section V we discuss a series of important limitations of our findings and several tests that suggest that these limitations do not drive our results. We conclude in Section VI with a discussion of future work and implications for the designers of interactive systems.

## II   BACKGROUND

*Generativity*

Because most research on online peer production has focused on the most successful projects (see Crowston et al., 2010), we still know very little about why some peer production efforts become highly generative while the vast majority never attract contributors. Al-though foundational theories in peer production offers guidance, we must first elaborate on these theories to formulate testable hypotheses about the antecedents of generativity.

Zittrain (2008) posits the "Principle of Procrastination" that proposes that generative technologies tend to be designed in a way that leaves most details for later saying, "generative systems are built on the notion that they are never fully complete, that they have many uses yet to be conceived of, and that the public can be trusted to invent and share good uses." For example, Zittrain suggests that the Internet was a more effective platform for innovation than corporate networks like *Prodigy* and *Compuserv* because its relative simplicity offered fewer constraints for potential innovators. In his influential essay, *The Cathedral and the Bazaar*, Raymond (1999) suggests that FLOSS projects like Linux attract participants because they "release early, release often" – that is, they publish their code earlier encouraging more collaboration in the form of feedback, bug fixes, and improvements.

Although there are several possible mechanisms through which "procrastination" and early releases might lead to generativity, one mechanism is the relative simplicity of these works. Early stage and incomplete projects will be simpler and easier for would-be contributors to understand and build off. Because these earlier, less complete, works are buggier, or more open-ended, they may also offer more avenues for engagement.

But while we interpret theory as suggesting that increased simplicity will be associated with generativity, this seems unlikely to hold for extremely simple works. The earliest possible release of Linux would, by definition, do nothing. It seems very unlikely that a featureless or extremely broken operating system kernel would excite and elicit contributions from other programmers in the way that Linux did. Similarly, if the designers of the Internet procrastinated on *all* features and created nothing, it seems very unlikely that their system would have been an even more generative platform.

> *Hypothesis 1A: There will exist an inverse-U-shaped relationship between a work's complexity and its generativity.*

Exposure to a work is, by definition, related to its generativity in that a work has to be seen to be remixed. Theorists have suggested that the relationship between popularity and remixing may run deeper. The antecedents of remixes, unlike some other forms of peer production, almost always have identifiable authors (Sinnreich, 2010). Lessig's (2008) key examples include music videos based on widely popular news footage and popular music and films. In Lessig's account, the act of remixing is often understood as a social statement of parody or critique. Jenkins (2008) documents how youth use fan fiction to create remixes of popular and culturally salient products and symbols.

Within particular communities, research has suggested that more popular individuals attract more remixers (Cheliotis and Yew, 2009). Using surveys and interviews with musicians, Sinnreich (2010) suggests that remixing is about creating explicit connections with previous, culturally salient, creators and that, "mash-ups are premised on the notion of recognizability and critique of pop culture," and that mash-up artists avoid rare vinyl samples in favor of popular songs to maintain these connections.

To the extent that remixing relies on cultural salience, we expect works of greater salience to be more generative. However, since popularity of the work itself might simply measure exposure, we should operationalize salience by looking to the prominence or "fame" of a work's creator while controlling for the exposure of the work in question. In other words, after having been viewed the same number of times, we expect a work by a more prominent creator to be more generative than a work by a less prominent author.

> *Hypothesis 1B: The prominence of a work's author will be positively related to its generativity.*

We suggest that a third determinant of generativity is "cumulativeness:" a term we borrow from Murray and O'Mahony (2007) to describe works that aggregate the efforts of many individuals through accretion and accumulation. Cheliotis and Yew (2009), Healy and Schussman (2003), and others have shown that activity in remixing communities is distributed unequally and that only a very small number of peer production projects incorporate the work of a large number of individuals building on one another's efforts. The majority of efforts are largely, or even entirely, uncollaborative.

Cheliotis and Yew (2009) have suggested that highly unequal rates of collaboration among projects in the ccMixter community is driven by a process of "preferential attachment" (Barabási and Albert, 1999) or cumulative advantage (DiPrete and Eirich, 2006) where, "works exhibiting a high degree of reuse become more attractive for further reuse." Cheliotis and Yew also suggest that, with important limits, remixing behavior will tend to form "chains" of remixed-remixes. To the extent that ccMixter is representative of other peer production projects in that collaboration drives more collaboration, we expect that cumulative remixes will be more generative than non-cumulative *de novo* works.

> *Hypothesis 1C: Works that are remixes themselves will be more generative than* de novo *projects.*

## Originality

Although theory on the relationships between remixed media, its creators, and the nature of their remixes is less developed, we find justification in existing theory for three hypotheses that parallel our hypotheses about generativity. In all three cases, we believe that theory points toward hypotheses suggesting that the qualities associated with higher generativity are also associated with lower originality in the resulting remixes.

In the previous section, we hypothesized that one possible mechanism behind "release early, release often" and the Principle of Procrastination is that simple projects are easier for new contributors to build on. Zittrain suggests that the generativity of a work will be determined, in part, by how easily new contributors can master it. Based on this, we posit that moderately simple works might be more generative than more complicated or simpler works because they are accessible to a relatively larger group of potential remixers. Driven by a marginal decrease in the effort or skill necessary to remix, we suggest that the remixes produced through this process will also involve less effort or skill and, as a result of these linked processes, will be relatively less transformative.

Although it is also possible that more complex works are closer to "completion" than relatively simpler works and are therefore subject to less intensive improvements, we suggest that originality in remixes will be driven primarily by wider participation in the act of remixing, and that, as a result, remixes of works of intermediate complexity will tend be less original than very simple or very complicated works.

> *Hypothesis 2A: There will exist a U-shaped relationship between a work's complexity and the originality of its derivatives.*

When discussing generativity above, we hypothesized that the creation of remixes of highly prominent creators is one way that remix artists seek cultural resonance for their works. To achieve this, it is important that a remixer maintain the recognizability of the original. For example, several musicians interviewed by Sinnreich suggest that P. Diddy's song *I'll be Missing You* became a cultural and commercial success in part because it consisted largely of a minimally modified version of the 1983 song *Every Breath You Take* by the band The Police.

Sinnreich argues that highly derivative remixes of culturally salient works strive to maintain a high degree of recognizability with the antecedent in the remix. To the extent that remixing of prominent work is more likely to be a form of cultural conversation, we will also expect the remixes of more popular or culturally salient works to be remixed lightly. On the other hand, when remixing the work of less prominent creators, the choice of a particular work might be driven more by use-value and, as a result, recognizability may play a less important role.

> Hypothesis 2B: The prominence a work's author will be negatively related to the originality of its derivatives.

Raymond (1999) describes "Linus' law" – "with enough eyeballs, all bugs are shallow" – to suggest that collaboratively produced software will be higher quality and less buggy. Benkler's theory of peer production suggests that it is lightly motivated individuals contributing small amounts who participate in some of the most collaborative, and most cumulative, works of peer production. As a result, we might expect cumulative remixes (i.e., remixes of remixes) to begin with a less buggy or more complete work and, as a result, have less work to do.

Also suggesting a limit to generativity, Cheliotis and Yew (2009) observe that when a project is very cumulative and the product of many subsequent reuses, it becomes less likely to be reused in future generations. To the extent that the "chain" network structure becomes decreasingly likely to continue as it grows in length, we might expect that the existence of a shared goal (stated or implicit) for cumulative work may influence its continuation. Although Cheliotis and Yew do not present data on the originality of remixes, one explanation of their observation on chain remixes is that cumulative remixing will, on average, represent a process of refining and elaborating that has limits.

> Hypothesis 2C: Remixes of works that are remixes themselves will be less original than remixes of *de novo* projects.

We do not suggest that these six hypotheses reflect a complete theory of generativity or originality in peer production communities. These hypotheses reflect our attempt at a partial theory in that they attempt to highlight three of the most widely cited theoretical determinants of generativity and originality. To our knowledge, none of these hypotheses has been tested empirically.

## III  EMPIRICAL SETTING AND METHODS

*Scratch*

To test our hypotheses, we turn to the Scratch online community: a public and free website with a large community of users who create, share, and remix interactive media. The community is built around the Scratch programming environment: a freely downloadable desktop application, akin to Adobe Flash, that allows amateur creators to combine images, music, and sound with programming code (Resnick et al., 2009). A screenshot of the Scratch programming environment is shown in Figure 1. Scratch was designed by the Lifelong Kindergarten Group at the MIT Media Lab as a platform for constructionist learning (e.g. Papert, 1980) and aims to introduce young people to computer programming. Scholars have located much of the practice and promise of remixing in communities of young technology users (e.g., Lessig, 2008; Jenkins, 2008; Palfrey and Gasser, 2008). With a large community of young users, Scratch represents an ideal platform to study remixing.

From within the Scratch authoring environment, creators can publish their projects on the Scratch community website hosted at MIT.[1] As of April 2012, more than one million users had created accounts on the website and more than one third of these users had shared at least one of more than 2.3 million total projects. As the only web community built around sharing Scratch projects, the community contains virtually all Scratch projects shared online.

The nature of Scratch projects varies widely and includes everything from interactive greeting cards to fractal simulations to animations to video games. The community is visited by more than half a million people each month[2] who can browse material on the website but visitors must create accounts in order to download projects or contribute in the form of publishing, commenting, showing support, tagging, or flagging projects as inappropriate. A majority of the community's users self-report their ages ranging between 8 to 17 years old with 13 being the median age for new accounts. Thirty-five percent of users of the online community self-report as female.

---

[1] http://scratch.mit.edu.
[2] http://quantcast.com/scratch.mit.edu

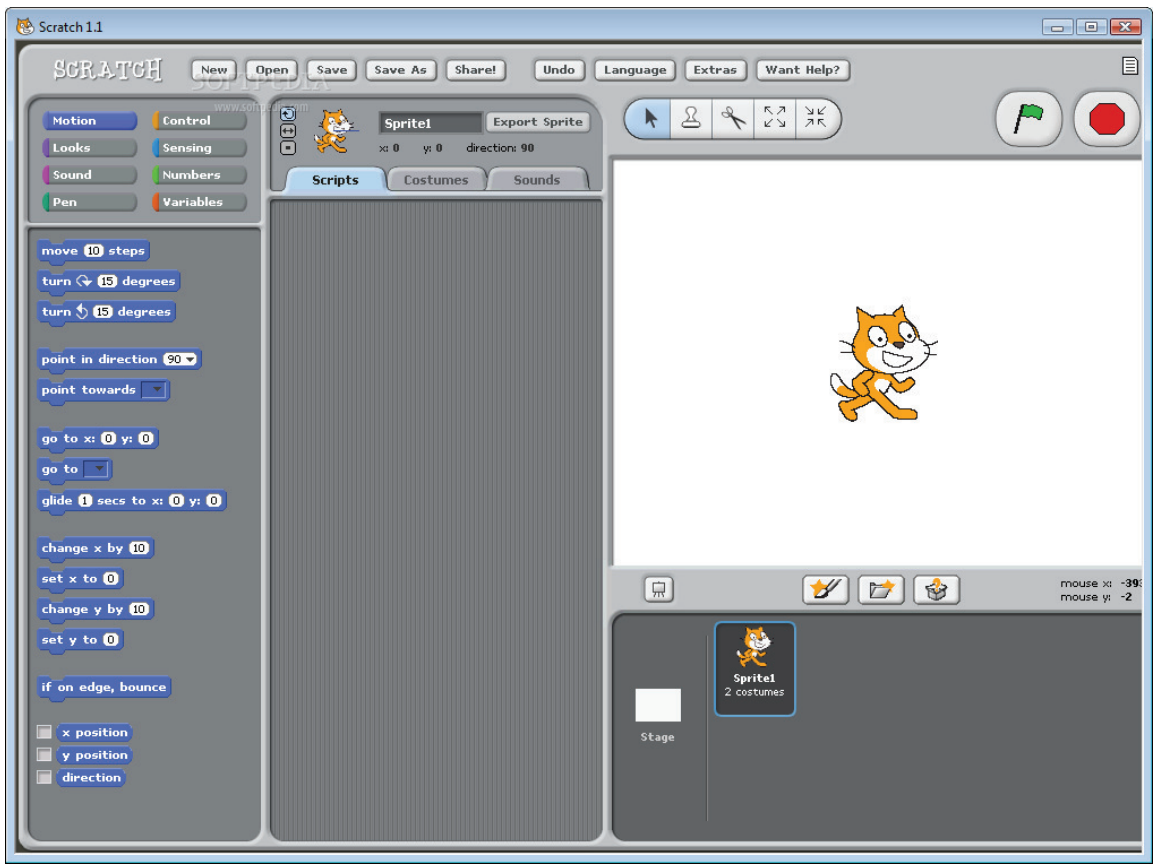

Figure 1: Screenshot of the Scratch desktop application. The leftmost column represents an inventory of possible programming blocks (shown in blue) which are assembled into program code in the center column. The white area in the top right represents the project as it will be displayed to a user interacting with the finished product. The bottom right column shows the available images (i.e., sprites) which are controlled by the code.

Central to the purposes of this study, the Scratch online community is designed as a platform for remixing. Influenced by theories of constructionist learning in communities (Papert, 1980) and communities of practice (Lave and Wenger, 1991), the community seeks to help users learn through exposure to, and engagement with one another's work. The commitment to remixing is deep and visible in Scratch. The name "Scratch" is a reference to hip hop disc jockeys' practice of remixing. Every project shared on Scratch is available for download and remix by any other user, through a prominent download button. Ad-ditionally, every project is licensed under a Creative Commons license – explained in "kid friendly terms" – that explicitly allows reuse. Administrators and community members routinely encourage remixing.[3]

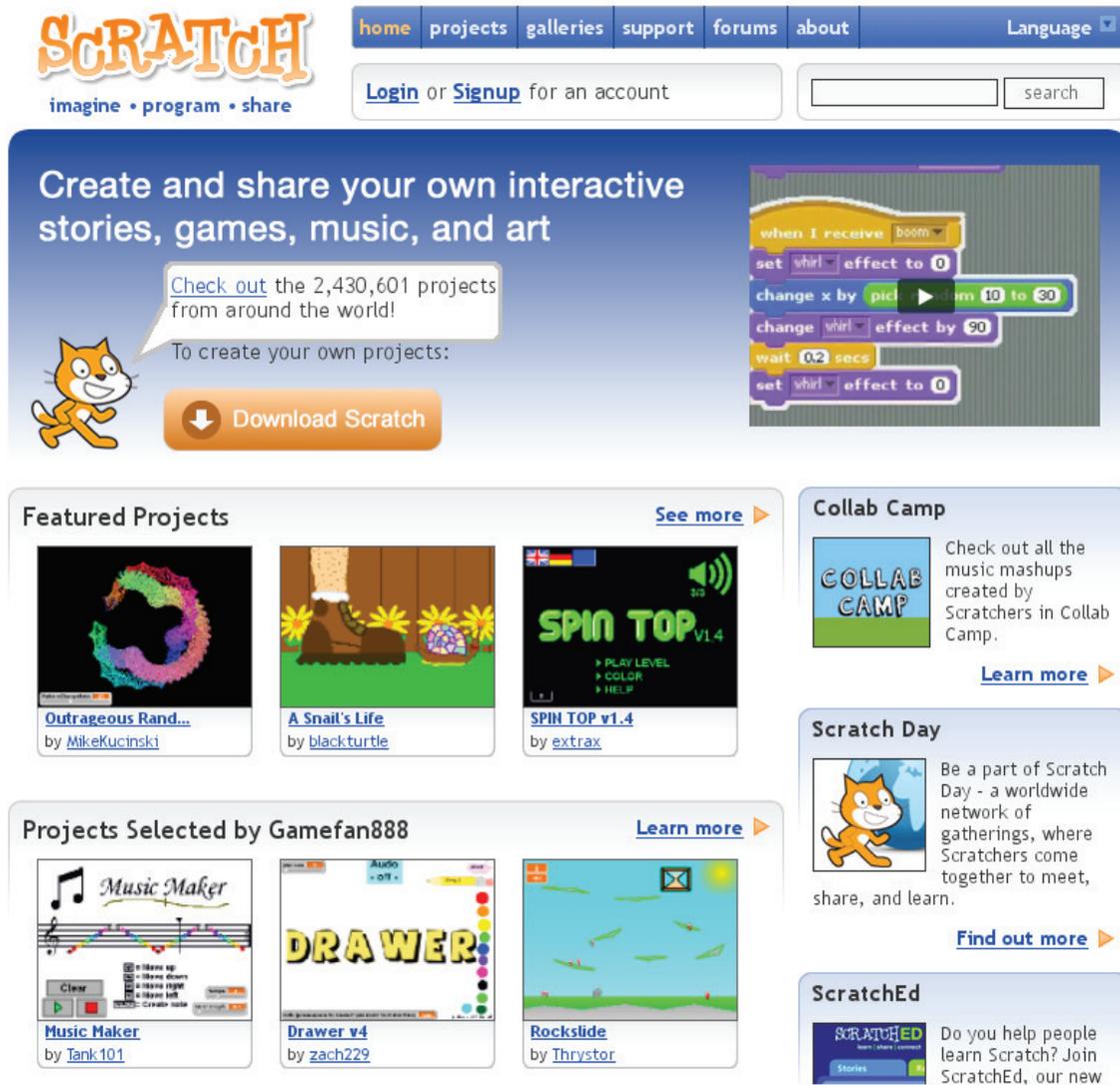

Figure 2: Screenshot of the Scratch online community front page as it appears to users visiting it from the Internet. The page is dynamically updated every several minutes to highlight new activity. (Accessed: April 1, 2012)

Issues of generativity and originality play a prominent role in the Scratch community. Although Scratch is designed as a remixing site, experience suggests that only around one tenth of all projects of Scratch projects are likely to be ever be remixed. Approximately 2% of remixes are flagged as inappropriate – often with accusations of unoriginality like, "This is MY own artwork he has uploaded without an ounce of originality." On the other hand, Scratch creators often explicitly encourage others to remix their works and even request help creating new features and solving bugs. Scratch users frequently respond to these requests with remixes but also frequently remix without prompting or communication.

*Data and Measures*

The Scratch online community is built on top of a large relational database that contains a wide variety of metadata on projects, users, and interactions on the website. Crucially, the Scratch website identifies, tracks, and presents data on whether and when projects are created via remixing. Additionally, the website stores each of the "raw" Scratch project files which can be further analyzed to reveal details, such as each project's programming code. Our dataset is constructed both by exporting metadata about the community's users, projects, and interactions, and by detailed algorithmic analyses of each project to compute the differences in code between remixes and their antecedents.

Our unit of analysis is each Scratch work or project ($p$). Our dataset consists of the 536,245 projects shared in the Scratch Community in a one-year period over 2010.[4] We do not include data on 1,182 projects which we were not able to analyze due to technical errors in our tools or corruption in the project files. We also omit 136,968 projects (21%) that were removed from the site by their creators, although we find, in robustness checks not reported here, that our results are not substantially different when we include them. The projects in our dataset were shared by 105,317 unique users ($u$). We selected data from 2010 because Scratch's administrators felt that the site and its community were mature and stable during this period and because the Scratch website did not undergo any significant design changes that might have affected remixing behavior. Because our data is longitudinal, we track each project for one year and, for time-varying measures, present measures at the end of the one year period.

---

[3] http://info.scratch.mit.edu/License_to_play
[4] Due to the timing of data collection, we use all the projects that were created during a one-year period from December 1, 2009 through December 1, 2010. We follow each project from this window for one year and our last data collected is from December 1, 2011.

| Variable | N | Mean | SD | Min | Max |
|---|---|---|---|---|---|
| **Dependent Variables** | | | | | |
| Remixes > 0 times in 1 yr. ($remixed_p$) | 536245 | 0.07 | 0.26 | 0.00 | 1.00 |
| Remixes within 1 yr. ($remixes_p$) | 536245 | 0.15 | 1.78 | 0.00 | 658.00 |
| Edit Distance (Mean) ($distance_p$) | 37512 | 85.57 | 397.66 | 0.00 | 21970.00 |
| **Question Predictors** | | | | | |
| Number of blocks ($blocks_p$) | 536245 | 99.60 | 476.19 | 0.00 | 196509.00 |
| User's cumulative views ($userviews_{up}$) | 536245 | 1563.59 | 5546.90 | 0.00 | 197844.00 |
| Remix status ($isremix_p$) | 536245 | 0.18 | 0.38 | 0.00 | 1.00 |
| **Controls** | | | | | |
| User age in years ($age_{up}$) | 523092 | 17.57 | 11.62 | 4.00 | 74.75 |
| Account age in months ($joined_{up}$) | 536245 | 4.79 | 7.18 | 0.00 | 45.43 |
| User is Female ($female_u$) | 536222 | 0.37 | 0.48 | 0.00 | 1.00 |
| Blocks per sprite ($blocks/sprites_p$) | 536245 | 11.82 | 22.75 | 0.00 | 3111.50 |
| Views within 1 yr. ($views_p$) | 536245 | 13.57 | 69.90 | 0.00 | 4977.00 |

Table 1: Summary statistics for variables used in our analysis. Measures with the subscript $p$ are measured at the level of the project while measures with the subscript $u$ are measured at the level of the user.

To answer our first three hypotheses, we operationalize generativity in two related ways. First, we construct a count of the number of remixes of each project shared within the first year of the project's publication ($remixes_p$). Second, we construct a dummy variable indicating if a project has been remixed by another user at least once in the one year period subsequent to being shared ($remixes_p > 0 \Rightarrow remixed_p = 1$). Because multiple-remixing is often endogenous (i.e., individuals may choose to remix a previous remixed project because others have remixed it), we suggest that our dichotomous measure ($remixed$) offers a more reliable, if more conservative, measure of generativity.

We operationalize complexity of projects as blocks ($blocks_p$). Blocks, shown in Figure 1, are analogous to tokens or symbols in source code for computer programs. Blocks are similar to, but more granular than, source lines of code which have a long history of use as a measure of both complexity and effort in software engineering (Walston and Felix, 1977; Albrecht and Gaffney, 1983). Scratch projects may also contain media elements. Several possible metrics of media complexity (e.g., the number of images or sounds in a project) are highly correlated with ($blocks_p$). For that reason, because an integrated code and media measure is not available, and because we feel that a code-based metric is more granular and comparable across projects, we leave exploration of media-based complexity as an area for future work.

Data on user interactions in Scratch provide a series of possible measures of user prominence within the community. Possible indicators of prominence include the total past views of a user's projects by other users, the number of past expressions of admiration or "loveits" (analogous to "liking" something on other social media platforms), total previous selections of a user's work as another's favorite, and total past downloads – each variable is measured at the level of the project or work but is aggregated for each project's creator at the point in time that a project is shared. Because these measures are highly correlated ($0.84 < \rho < 0.97$), we operationalize prominence as a user's cumulative previous views ($userviews_{up}$) at the moment that the project in question was uploaded. Our results are similar using the other indicators. We operationalize cumulativeness using a dummy variable that indicates whether a project itself is a remix of another project ($isremix_p$).

Finally, we include a series of control variables that may also be associated with the generativity of projects and with the originality of subsequent remixes. For each user, we include self-reported measures of gender which we have coded as a dichotomous variable ($female_u$), date of birth which we have coded as age in years at the moment that each project was shared ($age_{up}$) which may indicate sophistication of the user, and age of each account ($joined_{up}$) which may indicate the level of experience of a user with Scratch. For each project, we are concerned with the effect of exposure on the likelihood of remixing, so we attempt to control for views using the number of times that each project was visited in its first year on the site ($views_p$).[5] "Sprites" are the objects in Scratch project to which code is attached. Because more modular projects may be easier to remix, we also calculate a measure of the average numbers of blocks per sprite ($blocks/sprites_p$) which may act as a very coarse measure of modularity.

To answer Hypotheses 2A-C about originality, we create a new dataset that includes only the subset of 37,512 projects that were shared in the community during our one-year window and that were remixed at least once in the following year.[6] We operationalize the originality of a remix using a calculation of the degree to which a project diverges from its antecedent. To calculate this divergence, we begin with the list of remix-antecedent pairs. Next, we identify and compare each code component of the remix to the corresponding code component in the antecedent. Our measure of originality is the Levenshtein "edit distance" (Levenshtein, 1966).

---

[5]As an alternate specification, we instead control for the number of unique users who have views the site, with nearly indistinguishable results.

Levenshtein distance is a metric that has been used widely in software engineering to measure the divergence of code. The traditional Levenshtein analysis is a character-by-character comparison. In our case, we use blocks as tokens and our measure of distance is the sum of distances across all code and represents the minimum number of changes to blocks that would be needed to convert an antecedent project into its remix. This metric come with several important caveats. First, the measure will not capture artistic charges. If every media element in a project were changed but the code left intact, our analysis would consider the projects perfect copies. That said, we draw some confidence from the fact that exploratory analyses suggest that media derivativeness and code derivativeness are highly correlated. Additionally, the measure does not reflect "conceptual" remixing – such as employing Disney or Nintendo characters in a new Scratch game. We hope to address these limitations in future work.

Of course, a given project can be remixed multiple times. In fact, 11,704 of the projects remixed in our window (31%) were remixed more than once within a year of being shared. The distribution of remixes was highly skewed: the maximum number of remixes in our sample was as high as 658, and the mean was 2.14. As a result, our measure of edit distance is the mean edit distance of all projects shared in the year following a project's publication on the website ($distance_p$).

*Analysis*

Our analytic strategy involves the estimation of a series of two sets of parallel regression models. In both cases, we include variables operationalizing project complexity, creator prominence, and project cumulativeness that correspond to our three sets of parallel hypotheses. Both *blocks* and *userviews* are highly skewed but a started log transformation re-sults in an approximately normal distribution in each case. Hypothesis 1A and 2A predict a curvilinear relationship between the dependent variables and our measure of complexity. As a result, we include a quadratic specification for log *blocks* in each model and focus our interpretation on the coefficient associated with the quadratic term which will determine the direction of the curve. Because the amount of code in a remix does not reflect the work of only the person sharing the project, we include an interaction term between our measure of *blocks* and *isremix* to capture the difference in the effect of complexity between remixes and *de novo* projects.

---

[6]Due to technical errors or corrupted project files, we do not include 1,217 projects that site-metadata indicates were remixed but that we were unable to analyze. We believe that these errors were due to random corruption and are unlikely to bias our results.

Providing tests of Hypotheses 1A-C about generativity, our first two models consider generativity in the full dataset of 523,069 projects shared in our window of data collection for which we have complete information.[7] In our first and more conservative test, Model 1, we use logistic regression to model the likelihood of a project being remixed at least once on our sets of predictors and controls:

$$\text{logit}[\text{P}[remixed_p]] = \beta + \beta \log blocks_p + \beta \log blocks_p^2 + \beta \log userviews_{up} +$$
$$\beta isremix_p + \beta age_u + \beta joined_{up} + \beta female_u + \beta \log blocks/sprites_p +$$
$$\beta \log views_p + \beta(\log blocks_p \times isremix_p) + \beta(\log blocks_p^2 \times isremix_p)$$

Model 2 also tests Hypotheses 1A-C using our second measure of generativity: the count of remixes of each project in the first year. It is otherwise identical to Model 1. Poisson regression is frequently used for count dependent variables but, as is common with counts, there is an over-dispersion of zeros in the number of times a project has been remixed. To address this overdispersion, we use a negative binomial regression strategy that estimates the right side of the equation in the model above on the count of *remixes*.

To test Hypotheses 2A-C about originality, we begin with a reduced dataset that consists of the subset of 36,722 projects which were remixed at least once after being shared, and for which we have the creator's age and gender data. The right side of Model 3 is, once again, identical to that of Model 1 shown above. The left side of corresponds to the mean Levenshtein distance of every remix of the antecedent project. Because *distance* is a count and, like *remixes*, is overdispersed, we once again forgo Poisson regression in favor of a negative binomial count model.

---

[7]We omit 13,176 projects for which we are missing age or gender data.

# IV  RESULTS

Model 1 seems to provide support for Hypothesis 1A; we see support for the inverted U-shape in the relationship between complexity and generativity in the negative coefficient on the quadratic term in Models 1 and 2 (see Table 2). Holding other variables at their sample mean, Model 1 predicts that a bit more than 1 percent of projects will be remixed at both the minimum (0 blocks), and maximum (196,509 blocks) in our sample. That said, this support comes with a critical qualification. For the vast majority of projects, marginal increases in complexity are associated with increased generativity.

An example can serve to illustrate this point. The distribution of projects by *blocks* is highly skewed toward more simple projects with the median project having only 26 blocks. In other words, although the most simple and the most complicated projects are indeed at lower risk of being remixed than projects of median complexity (i.e., in an U-shaped relationship), we estimate that projects have an increasing likelihood of being remixed into the 95th percentile of complexity. Even among very complicated projects, the relationship is effectively flat. For example, holding all other predictors at their sample means, Model 1 estimates that 6.77% of projects with 385 blocks (the 95th percentile) would be remixed while effectively the same proportion of otherwise identical projects with 1,204 blocks (the 99th percentile) would be.

As predicted in Hypothesis 2A, we see some support for an inverse-U-shaped relationship between complexity and originality by the negative sign on the quadratic log *blocks* term in Model 3. That said, there is little evidence that simple projects are associated with increased originality in remixes because the first-order term is not significantly different from zero, therefore suggesting a curvilinear relationship where edit distance is monotonically increasing with complexity over the range of our data. Indeed, holding other variables at the sample mean, Model 3 estimates that a project with 3 blocks (10th percentile) will have an average edit distance of 21 changed blocks while an otherwise similar project with 211 blocks (90th percentile) would, on average, be associated with an mean edit distance of 81 changes.

Tests of Hypotheses 1B and 2B on the relationship between author prominence and generativity and originality are given in the coefficients estimates associated with *userviews*: the log-transformed count of the number of times that other users have viewed a project's creator's work in the past. We find support for Hypothesis 1B on the positive relation-

|  | Generativity | | Originality |
|  | Model 1 | Model 2 | Model 3 |
|  | (P[*remixed*]) | (*remixes*) | (*distance*) |
|---|---|---|---|
| (Intercept) | −5.070*** | −5.045*** | 2.437*** |
|  | (0.031) | (0.029) | (0.053) |
| log *blocks* | 0.525*** | 0.374*** | −0.028 |
|  | (0.016) | (0.016) | (0.027) |
| log *blocks*² | −0.037*** | −0.035*** | 0.052*** |
|  | (0.002) | (0.002) | (0.003) |
| log *userviews*$_{up}$ | 0.023*** | 0.002 | −0.041*** |
|  | (0.003) | (0.003) | (0.005) |
| *is.remix* | 0.786*** | 0.426*** | −1.035*** |
|  | (0.045) | (0.045) | (0.071) |
| *age* | 0.000 | 0.007*** | 0.006*** |
|  | (0.001) | (0.001) | (0.001) |
| *joined* | −0.006*** | −0.006*** | 0.005*** |
|  | (0.001) | (0.001) | (0.001) |
| *female* | −0.003 | 0.106*** | −0.348*** |
|  | (0.012) | (0.012) | (0.021) |
| log *blocks/sprites* | −0.517*** | −0.375*** | 0.289*** |
|  | (0.012) | (0.012) | (0.018) |
| log *views*$_p$ | 0.840*** | 1.028*** | 0.153*** |
|  | (0.006) | (0.006) | (0.009) |
| log *blocks* × *isremix* | 0.318*** | 0.303*** | 0.160*** |
|  | (0.025) | (0.026) | (0.040) |
| log *blocks*² × *isremix* | −0.045*** | −0.032*** | −0.002 |
|  | (0.003) | (0.003) | (0.005) |
| $\theta$ |  | 0.265*** | 0.301*** |
|  |  | (0.003) | (0.002) |
| *N* | 523069 | 523069 | 36722 |
| AIC | 219860.275 | 307810.178 | 313334.551 |
| BIC | 220396.313 | 308390.886 | 313777.130 |
| log *L* | −109882.137 | −153853.089 | −156615.276 |

Standard errors in parentheses; † significant at $p < .10$; * $p < .05$; ** $p < .01$; *** $p < .001$

Table 2: Model 1 is a logistic regression model of the likelihood of a project being remixed within one year. Model 2 is a negative binomial regression model of a count of the times a project will be remixed within a year. Both use the full dataset of projects ($N = 523,069$). Model 3 is a negative binomial regression model of a count of the mean edit distance for all projects remixed within a year of being shared ($N = 36,722$).

ship between author prominence and generativity. Holding other variables at their sample mean, Model 1 predicts that the odds of being remixed are slighter higher (1.02 times) for each log unit increase in the number of previous loveits the project's creator received, and that the result is statistically significant. Model 2 adds an important qualification to this support. There is no statistically significant association with prominence when operationalized as the number of remixes within one year. In other words, author prominence is a positive predictor of whether or not a project will be remixed but is not associated with higher numbers of total remixes.

We also find strong support for Hypothesis 2B that predicts a negative relationship between creator prominence and originality of remixes. Holding other predictors at their sample mean, we estimate that remixes of a project whose creator's previous projects had received no previous "views" (10th percentile) would have an average edit distance of 44 changed blocks. An otherwise identical project whose creator had received 3,652 previous "views" (90th percentile) would be estimated to have an average edit distance of 32 changed blocks.

Our results also provide strong support for Hypotheses 1C and 2C. Tests of the association between cumulativeness and measures of originality and generativity are captured by the parameter estimates associated with *isremix*. Model 1 suggests that the odds that a remix will be remixed is 2.2 times higher than the odds that an otherwise similar *de novo* project will be. Model 2 suggests that remixes will also be remixed more times. In strong support of Hypothesis 2C, Model 3 suggests that these remixed remixes will tend to be much less original. Holding other qualities at their sample means, Model 3 estimates that a remixed remix will have an edit distance of 24 changed blocks while a similar *de novo* project will have an edit distance of 41 changed blocks.

The models also include a statistically significant parameter estimate associated with the interaction between complexity and cumulativeness as measured in blocks. Because our measure of blocks is non-linear, interpretation of this result is complex. Prototypical plots of the estimates for remixes and non-remixes across our sample's range of project complexity are shown in Figure 3. Both generativity models suggest that remixes are associated with higher rates of generativity for all projects in our dataset. Both models also suggest that the inverse-U predicted in Hypothesis 1A is likely stronger for remixes than *de novo* projects. Model 3 predicts that for nearly all projects in our sample, remixes of remixes will be less original, as measured by edit distance, than otherwise similar remixes

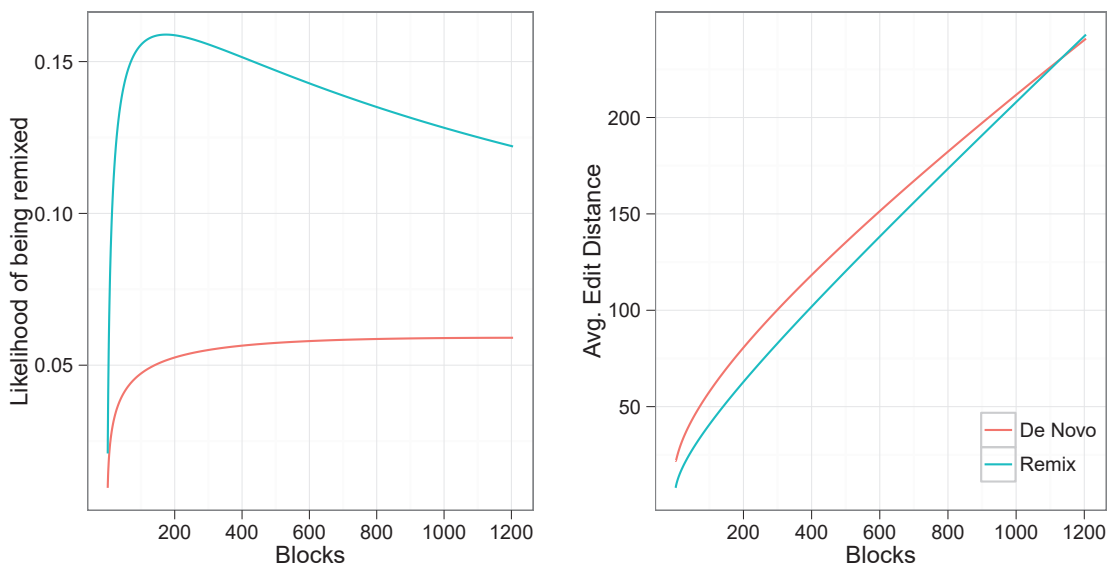

Figure 3: Two plots of estimated values for prototypical projects. Panel 1 (left) display predicted probabilities of being remixed as estimated by Model 1. Panel 2 (right) display predicted edit distances as estimated by from Model 3. Both models show predicted values for both remixes and *de novo* projects from 0 to 1,204 blocks ($99^{th}$ percentile).

of *de novo* projects, but that this difference is unlikely to be substantively meaningful.

## V    LIMITATIONS

There are a number of important limitations and threats to the validity of the results presented above. First, although our data is longitudinal, our analytic strategy follows each project for one year and treats these data as cross-sectional. Model 1 treats projects as being remixed only if they are remixed within one year of being published. Of course, projects can and are remixed after one year and there is a risk that our results are biased by the fact that these "late bloomers" systematically differ from other projects. We can model this threat using non-parametric Kaplan-Meyer survival functions of projects' likelihood of being remixed on a full dataset of Scratch projects. The resulting estimates suggest a rapidly flattening survival function. For example, 30 days after being shared, 8% of projects are remixed while after a full year, 10.8% are; only 2.4% more projects will have been remixed two years after that. As a result, we are confident that the analysis presented considers a large majority of remixing activity. Of course, this does not rule out the threat that these "late bloomer" projects are very unusual and our results may be biased through their omission. In other analyses not reported here, we use Cox proportional hazard survival and counting process models to estimate the instantaneous risk of projects being remixed using a random subset of 100,000 projects. The signs and relative magnitudes of the coefficients in our model are not substantively different from those presented here.

A second concern is that the use of average edit distance in Model 3 may lead us to conclude that more generative projects will tend to have lower edit distances simply because all projects are susceptible to unoriginal remixing and that being remixed more often puts projects at increased risk of these very simple remixes. We address this threat by re-estimating Model 3 using the highest edit distance of any remix as our measure of originality. The results of this model are largely unchanged from those reported in Table 2 and, indeed, are even stronger in the estimates of the effect of prominence and cumulativeness.

Third, there is an important concern with blocks as an indicator of complexity. As we have already suggested, blocks will not capture complexity in ways that do not involve programming, such as story-telling and visual arts. They can also be "cut and pasted" in a way that may not correspond to complexity through increased effort. It is possible that very complicated cut-and-paste projects are skewing the results for complexity. We can address this with a unique measure available in Scratch. Each Scratch project records the time and date every time that a user clicks the "save" button as well as the time that the user shares the project. We can use the time between the first "save" and the point at which the user shared the project as a proxy for effort.

This alternative measure is noisy in the sense that some users may not share a project for hours, days, or weeks, but not spend that entire period engaged in work on the project. Additionally, 44% of the projects in our window were shared without ever being saved once, so values on this indicator are missing. With these limitations in mind, we re-estimate Models 1 and 2 on the subset of 298,926 projects for which we have data and replace our measure of blocks with "minutes-to-share" (*MTS*). We find that our results in modeling generativity using this alternative specification are essentially unchanged. A similar re-estimation of Model 3 using the 22,048 remixed projects with *MTS* data did not find support for either the quadratic specification of *MTS* or its interaction with *isremix*, but offered substantively similar predictions in its estimation of a positive linear association between edit distance and complexity, which leaves our findings essentially unchanged. This robustness check also give us additional confidence in the applicability of these results to media and other non-code complexity.

In other robustness checks, we add random effects to control for possible clustering due to the fact that a single user can upload multiple projects and find that our estimates and results are unaffected. We also use robust estimation of standard errors to address concerns of potential heteroscedasticity. Using robust estimates for Model 3, the interaction terms between *blocks* and *isremix*, already substantively similar, are rendered statistically insignificant. The rest of our results, and our findings for each of our hypotheses, are unchanged.

A final concern, common to studies of peer production, is the question of generalizability. We have tested generalizability to other periods of time within the life of the Scratch community and find that our results are similar. Although we cannot speak to the generalizability of these results to other remixing communities or peer production projects, we believe that remixing in Scratch provides insight into the behavior of young creators more generally. The degree to which these results will generalize to adults, to other communities, or to activities other than the creation of animations and games, remain largely open questions for future research.

## VI  DISCUSSION

This paper provides support for the following paradox: attributes of works associated with increased generativity are associated with decreased originality, and vice versa. Our findings are based on a set of six hypotheses built on foundational theories of peer production which are tested using data from the Scratch online community. We find at least some support for the hypotheses that a work's generativity has a U-shaped relationship with its complexity (H1A), that it is positively related to the prominence of the work's creator (H1B), and that it is positively related to the work's cumulative nature (H1C). We also find at least some support for the hypotheses that the originality of remixes will have an inverse-U-shaped relationship to the complexity of the antecedent work (H2A), that it is negatively associated with the prominence of the antecedent's creator (H2B), and that it is negatively related to the work's cumulative nature (H2C).

This paper's primary contribution for system design theory is the proposal of a critical trade-off between the quality and quantity of remixes. To the extent that these results generalize, designers may need to trade-off deeper remixing with increased collaboration. Our findings imply difficult decisions around manipulating the visibility of variables such as au-thor prominence and project complexity. For example, designers of a new peer-production system in need of more content might want to build features that further emphasize the salience of author prominence and remix "chains" in order to encourage generative content. However, our findings suggest that these designs might come at a cost in terms of the originality of the derivative works. Our results regarding the relationship of complexity to generativity and originality of remixes suggest that supporting increased complexity, at least for most projects, may have fewer drawbacks.

Many social media sites, including YouTube and DeviantArt, track and display user prominence using a metric of aggregate views nearly identical to our operationalization of prominence. Other sites try to incentivize collaboration with prominence through leaderboards. Our results suggest this technique can lead to increased generativity but might also lead to a decrease in originality due to the incentive itself. We also suggest that it may be important to avoid rewarding correlates of generativity for their own sake when it is generativity that a designer wants to encourage. Encouraging cumulativeness by incentivizing or raising the visibility of cumulative projects may be another way for system designers to encourage generativity; but, once again, our results suggest it may also be at the expense of originality of the resulting remixing.

Surprisingly, our weakest support is for the hypotheses about complexity that stem from our elaboration of Zittrain's "Principle of Procrastination" and Raymond's exhortation to "release early and release often" – the most widely cited theories of generativity. We find support for our hypothesis that the most complex projects will be less generative than projects of moderate complexity, but only if we consider the very most complex examples. In general, we find largely positive relationships between complexity and both generativity and originality over most of our data. This may point in the direction of one potential solution to the "remixing dilemma" we propose. It may also be that the young users of Scratch are unlikely to create projects that are complex enough to trigger the effect suggested by theory. It may also be that complexity is simply a poor measure of completeness, earliness, or open-endedness as it is theorized by Zittrain and Raymond. More research is needed to clarify this relationship.

It also bears noting that designers seeking implications in our research are often working at the level of socio-technical systems while our investigation is focused on the qualities of content shared within a single system. By holding the system constant, we hope to offer a deeper understanding of social dynamics that is essential to the design of well-functioning systems. That said, our findings are no substitute for experimental validation with between-community or longitudinal experimental designs. We see these as rich areas for future research.

Of course, nothing we have shown devalues the promise of remixing in terms of peer production, culture, and innovation. Indeed, we believe that societies' ability to harness the power of remixing is deeply important, but requires further analyses similar to ours. Though our results suggest that highly generative works that lead to highly original derivatives may be rare and difficult for system designers to support, we do not suggest that encouraging them is anything but a worthwhile, and critically important, goal.